\documentclass[preprint,showpacs,preprintnumbers,amsmath,amssymb, aps]{revtex4}

\usepackage{verbatim}
\usepackage{graphicx}
\usepackage{dcolumn}
\usepackage{bm}
\def\coeff#1#2{{\textstyle {\frac {#1}{#2}}}}
\def\half{\coeff 12}

\def\N{{\cal N}}

\def\S_1{{\widetilde {S_1}}}

\def\R{{\mathbb R}}

\def\tr{{\rm tr}}

\def\Z{{\mathbb Z}}

\def\Dslash{{\rlap{\raise 1pt \hbox{$\>/$}}D}}

\begin{document}

\preprint{SLAC-PUB-13604}
 
\title{ Chiral gauge dynamics and  dynamical supersymmetry breaking
\\ }

\author{ Erich Poppitz} 
 \affiliation{Department of Physics, University of Toronto, Toronto, ON M5S 1A7, Canada}
\author{Mithat \"Unsal}
 \affiliation{SLAC and Physics Department, Stanford University, Stanford, CA 94025/94305, USA}
 
\date{\today} 

\begin{abstract}
We study the dynamics of a chiral $SU(2)$ gauge theory with a Weyl fermion in the $I=3/2$ representation and of its supersymmetric generalization.  
 In the former,  
 we find a new and exotic mechanism of confinement, induced by topological excitations that we refer to as  magnetic quintets.  
  The supersymmetric version was examined earlier  in the context of dynamical supersymmetry breaking by Intriligator, Seiberg, and Shenker, who showed that if this gauge theory confines at the origin of moduli space, one may break supersymmetry by adding a  tree level 
 superpotential. We examine the dynamics by deforming the theory on  $S^1 \times \R^3$, and show that   the infrared behavior of this theory is   an interacting  CFT at small $S^1$. 
 We argue that this continues to hold at large $S^1$,  and if so, 
 that supersymmetry must remain unbroken. Our methods also provide  the  microscopic origin of various superpotentials in SQCD on $S^1 \times \R^3$---which were previously obtained by using symmetry and holomorphy---and  resolve  a long standing interpretational puzzle concerning a    flux  operator discovered by Affleck, Harvey, and Witten. It is generated by a topological excitation, a ``magnetic bion", whose stability is due to fermion pair exchange between its constituents.
 We also briefly comment on composite monopole operators as leading
effects in two dimensional anti-ferromagnets.

\end{abstract}

\maketitle

\tableofcontents

\section{Chiral $\mathbf{SU(2)}$ with $\mathbf{I={3\over 2}}$ fermion} 

Consider the $SU(2)$ Yang-Mills  theory with a single left handed fermion in the three-index symmetric ($I=3/2$)  representation of the gauge group. This theory is asymptotically free. Since  the fermion is in a half-integer (pseudo-real) representation, a gauge invariant fermion bilinear  
vanishes identically and 
 the theory is chiral.  The index of an  instanton is even,  ${\cal I}_{\rm inst}= 10$, and the  theory does  not suffer from the global (Witten) anomaly \cite{Witten:1982fp}.  The ${\cal{N}}=1$ supersymmetric version of this theory, if it  is confining,  provides the simplest  (in terms of  rank and matter content)  example of dynamical supersymmetry breaking \cite{Intriligator:1994rx}. However, the confinement hypothesis  in this theory  remains  controversial to date 
 \cite{Intriligator:1994rx, Shifman:1999mv,Intriligator:2005if}. Not much is known about the dynamics of the  non-supersymmetric version   as well.  In this work, we wish to study the dynamics of both theories  using recent techniques developed in \cite{Shifman:2008cx} by M. Shifman and one of us (M.\" U.). 
 We will argue that the non-supersymmetric theory confines via a new and very exotic mechanism, and  the supersymmetric theory does not confine. This implies that this chiral supersymmetric theory does not provide an example of   dynamical supersymmetry breaking.

Let us denote the $I\:$=$\:3/2$  fermion  in the   ${(1/2, 0)}$ representation of the Lorentz group 
$SU(2)_L \times SU(2)_R$ 
as $\psi_{\alpha, a bc}$ where $\alpha$ is a Lorentz index and the fermion is fully symmetric in the gauge indices $a,b, c$, i.e.~$\psi_{\alpha, a bc}\:$=$\:\psi_{\alpha, b a c}\:$=$ \ldots$. 
Consequently, $ \psi^2 =   
 \epsilon^{\alpha_1 \alpha_2}
\epsilon^{a_1 a_2}  \epsilon^{b_1 b_2}  \epsilon^{c_1 c_2}  
\psi_{\alpha_1,  a_1 b_1 c_1}  \psi_{\alpha_2,  a_2 b_2 c_2}  = 0$ and the leading non-vanishing multi-fermion operators are  $\psi^4 ,  \psi^6 $, {\it etc}.   
The theory has a classical  $U(1)$  symmetry, $ \psi \rightarrow e^{i \alpha}  \psi$. Quantum mechanically, it is reduced to  ${\mathbb Z}_{10}$ due to the instanton vertex:
 \begin{equation}
 \label{instanton1}
 I (x) = e^{ - S_{\rm  inst}} \psi^{10} \equiv  e^{ - \frac{8 \pi^2}{g^2}} \psi^{10}~ .
 \end{equation}
The action of    ${\mathbb Z}_{10}$  on fermions is:
  \begin{equation}
 {\mathbb Z}_{10}: \psi \rightarrow e^{i \frac{2 \pi k}{10}   }\psi,  \qquad k=0, \ldots   9.
 \label{Z10}
\end{equation}
A ${\mathbb Z}_{2}$ subgroup of ${\mathbb Z}_{10}$ is fermion number modulo two, $(-1)^F$, and cannot be spontaneously broken so long as Lorentz symmetry is unbroken.  An order parameter which may probe the chiral symmetry realization is $\psi^4$. Since the greatest common divisor ${ \rm gcd}{(10,4)}=2$,  $\langle   \psi^4   \rangle \neq 0 $ implies the chiral symmetry breaking pattern  ${\mathbb Z}_{10} \rightarrow  {\mathbb Z}_{2}$  and the presence of five isolated vacua. 
This is  the current state of knowledge about  this theory. We wish to understand the chiral dynamics by using the  techniques developed in  \cite{Shifman:2008cx}. 

\subsection{Deformation theory at work for a non-supersymmetric chiral theory}
\begin{figure}[ht]
\begin{center}
\includegraphics[angle=0, width=0.5\textwidth]{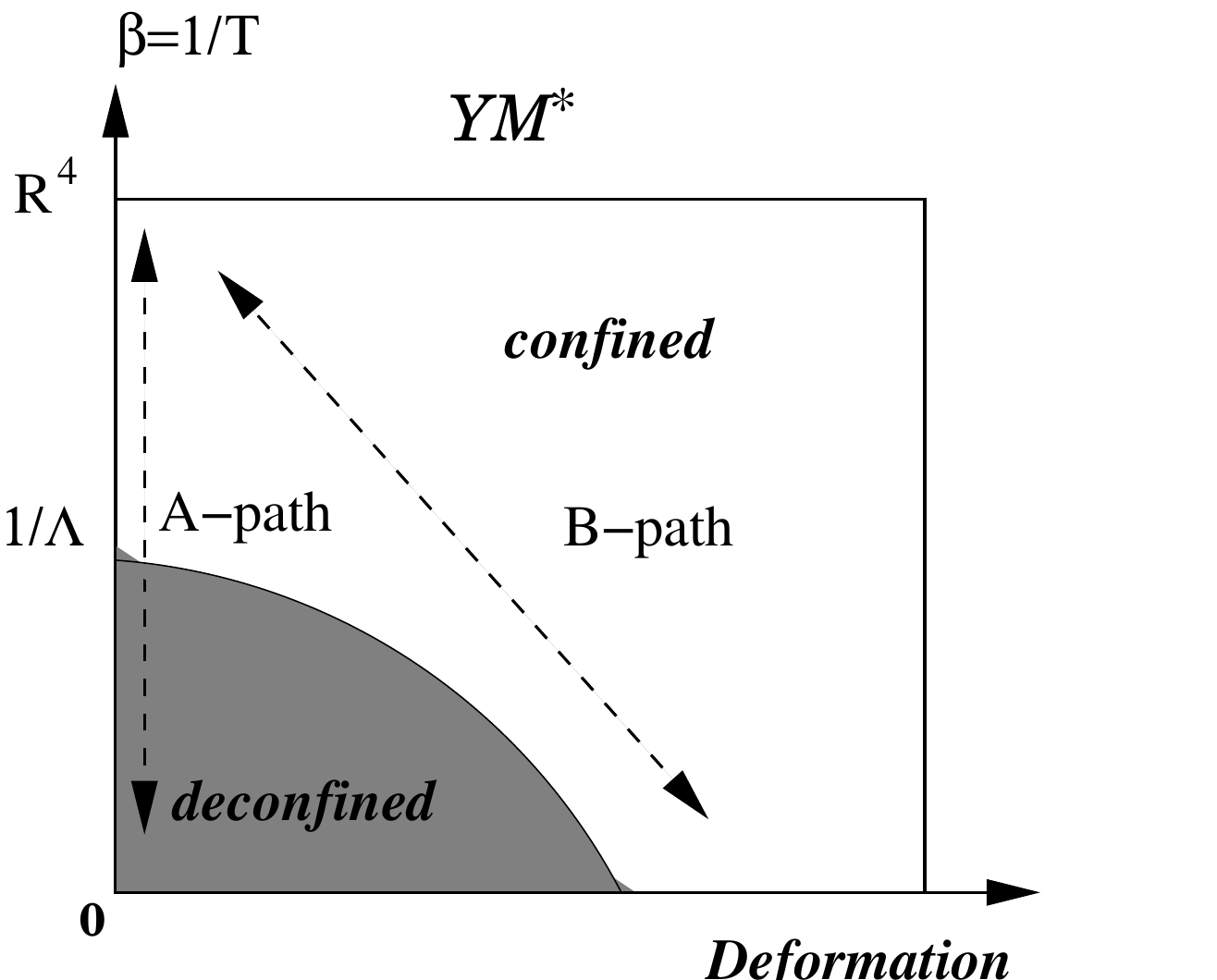}
\caption{ The figure describes the main idea of deformation theory. The A-path (split from y-axis for visual convenience) corresponds to pure YM theory. This theory undergoes a center symmetry changing transition when the radius is of order the inverse strong scale.  
The double-trace deformation  can be used to stabilize  the center symmetry down to arbitrarily small radius, along the B-path.  The deformed YM theory (YM*) is continuously connected to pure YM on $\R^4$.  The small-volume confined theory is amenable to non-perturbative semi-classical analysis, like supersymmetric theories.   The dynamics of YM theories with 
  vector-like  or chiral fermions can  be studied within this  framework.  
 }
  \label {fig:deform}
\end{center}
\end{figure}

We consider the  theory on a small $ S^1 \times  {\mathbb R}^3 $ and apply the double-trace deformations to  generate a repulsion between the eigenvalues of the Wilson line. 
This is a center-symmetry-stabilizing deformation in the pure YM theory.  In theories with fermions, the deformed gauge theories  without continuous non-abelian flavor  symmetries 
are conjectured to be smoothly connected to 
the {\it undeformed} theories on  ${\mathbb R}^4$. The  main idea is shown in Fig.\ref{fig:deform} and the  details are discussed in     \cite{Shifman:2008cx} and references therein.  
  This conjecture is  numerically shown to hold on the  lattice for the deformed YM theory \cite{Myers:2007vc}, and recent lattice simulations also confirmed the existence of center-symmetric  small-$S^1$ phases in gauge theories which reduce to deformed models for small $S^1$  \cite{Cossu:2009sq}. 
 Upon deformation, the
center symmetry on small and large $S^1$ is realized in the same way. 
 For example, in theories with   one fermion flavor  (in an arbitrary representation), one can analytically  show that the (discrete) chiral symmetry spontaneously breaks on small $S^1$, which is the 
 expected   behavior  on $\R^4$.   
One can also demonstrate mass gap and confinement by using abelian duality, within the region of applicability of semi-classical techniques. This is the basis for the  smoothness conjecture \cite{Shifman:2008cx}.  In the supersymmetric version of this theory, the deformation is not needed. 

In the small-$S^1$ regime of the deformed chiral $I=3/2 $ theory, the holonomy of the Wilson line behaves as an adjoint Higgs field, with two eigenvalues located at antipodal points, $\pm \pi/2$. 
The gauge structure at long distances reduces to an abelian gauge theory and   
 the infrared physics  can be described in terms of the perturbatively massless degrees of freedom. 
Due to ``gauge symmetry breaking", $SU(2) \rightarrow U(1)$, via a {\it compact} adjoint Higgs scalar,   there are two types of monopoles that we refer to as  BPS  $({\cal M}_1)$ and KK  $({\cal M}_2)$.  The number of fermionic zero modes for these two topological excitations (${\cal I}_1, {\cal I}_2$ for  ${\cal M}_1, {\cal M}_2$, respectively) can be extracted from the index theorem on $S^1 \times \R^3$ \cite{Nye:2000eg, Poppitz:2008hr}:
\begin{equation}
{\cal I}_{1} =4, \qquad   {\cal I}_{2} =6, \qquad    {\cal I}_{\rm inst}= {\cal I}_{1} + {\cal I}_{2} =10.
\end{equation}
 The corresponding (anti-)monopole operators are:
   \begin{eqnarray}
 &&  {\cal M}_{\rm 1} = e^{-S_0} e^{i \sigma} \psi^4, \qquad  \overline {\cal M}_{ 1} = 
   e^{-S_0} e^{- i \sigma} {\bar \psi}^4,  \cr \cr 
&&    {\cal M}_{\rm 2} = e^{-S_0} e^{- i \sigma} \psi^6, \qquad 
     \overline {\cal M}_{2 } =  e^{-S_0} e^{ i \sigma} {\bar \psi}^6,  
     \label{monopoles}
   \end{eqnarray}
   where  $ S_0 =  \frac{8 \pi^2} {N g^2} =  \frac{4 \pi^2} { g^2} $
 is the monopole action in the center-symmetric background and $d\sigma = *F$ is the dual photon (for brevity, we set  the numerical constants and couplings  appearing in the duality relation to unity).  The product of the BPS and KK monopole operators has the  quantum numbers of the instanton  (\ref{instanton1}):
   \begin{eqnarray}
   \label{inst-mon}
 I(x) \sim  {\cal M}_{\rm 1}   {\cal M}_{\rm 2}  \sim e^{-   S_{\rm inst} }  \; \psi^{10}, \qquad 
 S_{\rm inst} =    2 S_0 ~.
    \end{eqnarray}

In the absence of topological flux operators (which get induced by monopoles), the dual of the free Maxwell theory  enjoys a   $U(1)_J$  topological  shift symmetry: 
\begin{equation} 
U(1)_J: \sigma \longrightarrow \sigma + \alpha 
\label{u1j}
\end{equation}
which protects the masslessness of the dual scalar $\sigma$. The current associated with this symmetry is
${\cal J}_{\mu} = \partial_{\mu}  \sigma= \half \epsilon_{\mu \nu \rho} F_{\nu \rho} = F_{\mu}$ where $F_{\mu}$ is magnetic field. The conservation of $U(1)_J$,  $\partial_{\mu} {\cal J}_{\mu} = \partial_{\mu}  F_{\mu}=0$, is equivalent to the absence of monopole operators.  It should be noted that $U(1)_J$ is not a microscopic symmetry of the theory, as it does not act on the microscopic fields in any naive way, and it only emerges upon duality and should be viewed as an infrared symmetry. Throughout the paper, the topological symmetry and its discrete subgroups 
will play a major role in the construction of long distance theories. 

Clearly, because of fermion zero modes, neither the elementary monopoles, nor the instanton term provide  a mass term for the dual photon. Let us first demonstrate that a mass term for the photon is allowed by symmetries.
Since  ${\mathbb Z}_{10}$ of eqn.~(\ref{Z10})  is a true symmetry of the microscopic theory, it must also be a symmetry of the long distance theory. Were it not for the the topological $U(1)_J$ symmetry, 
the monopole operators (\ref{monopoles}) would have implied that  ${\mathbb Z}_{10}$ is anomalous, which is not correct. The fact that   ${\mathbb Z}_{10}$ is non-anomalous 
demands the intertwining of the microscopic ${\mathbb Z}_{10}$ symmetry with a compensator subgroup of the $U(1)_J$ such that the  monopole operator 
$ {\cal M}_{\rm 1} $ remains invariant.  For the example at hand,  
   \begin{eqnarray}
 {\mathbb Z}_{10}: \qquad  && \psi^4 \rightarrow e^{i \frac{8 \pi k}{10}   }\psi^4,   
 \qquad   (\Z_5)_J \subset U(1)_J : \qquad \sigma \rightarrow \sigma -   \frac{4 \pi k }{5}    ~.
 \label{photonshift}
    \end{eqnarray}
Since $\sigma$ is periodic by $2 \pi$, $k \sim  k+5$ are identified. Moreover,  the true action of    
 ${\mathbb Z}_{10}$  on the chiral order parameter $\psi^4$ is  $({\mathbb Z}_{5})_A$. Thus,  the long distance effective theory must rely on the linear combination of local and topological symmetry  
that we refer to as  $(\Z_5)_*$:
 \begin{equation}
(\Z_5)_A \times  (\Z_5)_J \supset (\Z_5)_*
\end{equation}  
$(\Z_5)_*$  acts on original  fields of the Lagrangian  as $(\Z_5)_A $, and on topological operators such as  $e^{i \sigma} $,   it acts as   $(\Z_5)_J$. 
  Note that  the KK-monopole operator $ {\cal M}_{\rm 2} $ is automatically  invariant under the 
 $({\mathbb Z}_{5})_{*}$  discrete shift symmetry. Such intertwining of microscopic and macroscopic (topological) symmetries  of fermions and the dual photon is generic in the presence of topological excitations on $\R^3$  and  $\R^3 \times S^1$, and is one of the powerful tools that we use throughout.

 The   $({\mathbb Z}_{5})_*$  discrete shift symmetry cannot prohibit a mass term for the dual photon, but   can delay it in an $e^{-S_0}$ expansion. 
In particular,  it  forbids all pure flux operators of the type $e^{i n \sigma}$ 
but  $(e^{i 5 \sigma})^{l}$ with an integer $l$.   Thus, the leading  pure-flux operator   appears at order $e^{-5S_0}$ in the topological expansion and is of the form: 
\begin{eqnarray}
\label{quintetpot}
 e^{-5S_0} ( e^{i5 \sigma}  +   e^{-i5 \sigma}  )  \sim  e^{-5S_0} \cos 5 \sigma~.
\end{eqnarray}
This is the first term in the semi-classical expansion  which is purely bosonic and, hence, 
can generate a mass gap in the 
gauge sector of the theory.  The dual photon mass is $m_{
\sigma} \sim \frac{1}{L}  e^{-5S_0/2}$, where $L$ is the circumference of $S^1$.
Using the one loop result for the renormalization group $\beta$ function:
\begin{eqnarray}
e^{-8 \pi^2/g^2} = (\Lambda L)^{\beta_0}, \qquad \beta_0= \frac{11}{3}
N - \frac{2}{3} T(j) N_f^{w}, \qquad T(j) = \frac{1}{3} j(j+1)(2j+1)\;
,
\end{eqnarray}
with $j=3/2$ and setting the number of Weyl spinors $N_f^{w}=1$,  we obtain
$m_{\sigma} \approx \Lambda(  \Lambda L)^4$ in the $\Lambda L \ll1$ domain.

Since $\sigma \sim \sigma + 2 \pi$, the potential (\ref{quintetpot}) has five isolated 
minima within the  fundamental domain. This  implies  spontaneous breaking of the       
$ {\mathbb Z}_{5}$ down to $ {\mathbb Z}_{1}$.  The  minima are located at: 
\begin{equation}
{\sigma_0} |_q= \frac{2 \pi} {5} q, \qquad  q=0, \ldots 4. 
 \end{equation} 
 In a Hilbert space interpretation, let us label  these vacua as $ |\Omega_q \rangle$. 
  Since the shift symmetry of the photon is intertwined with the discrete chiral symmetry in the small-$S_1$ regime,  this is the same as  spontaneous breaking of the discrete chiral symmetry. 
  Expanding the $\sigma$ field around the minimum, it is clear that there is a fermion condensate 
$\langle \psi^4 \rangle $ determined by the choice of the vacuum.  In particular, 
\begin{equation}
 \langle \Omega_q  |   \psi^4 |   \Omega_q   \rangle \sim  e^{-S_0} e^{i\frac{2 \pi} {5} q}, 
 \qquad  q=0, \ldots 4~. 
 \end{equation} 
which is  the expected ${\mathbb Z}_{10} \rightarrow  {\mathbb Z}_{2}$ pattern of the gauge theory on $\R^4$.

\subsection{Magnetic quintet}

\begin{figure}[ht]
\begin{center}
\includegraphics[angle=0, width=0.3\textwidth]{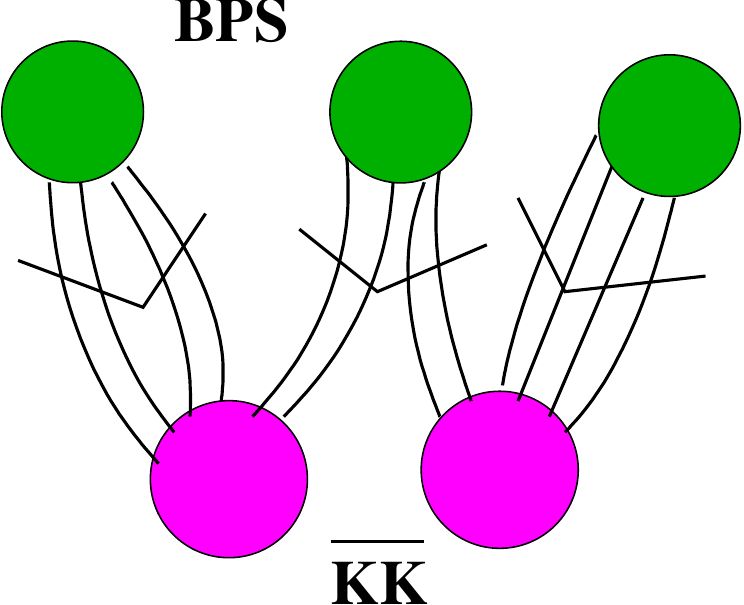}
\caption{
A cartoon of the magnetic quintet in the chiral non-supersymmetric $SU(2)$ gauge theory with an $I=3/2$ representation fermion. It may be viewed as a composite of 3 BPS and 2 $\overline {\rm KK}$ monopoles.  These excitations, all with magnetic charge +1,  repel each other in the absence of fermionic zero modes. The fermion zero mode exchanges generate a five-body interaction which leads to the formation of the magnetic quintet. 
This is the leading cause of the mass gap and the bosonic operator allowed by the $(\Z_5)_{*}$ symmetry. 
 An analogous topological excitation is forbidden in the supersymmetric theory by a continuous 
$U(1)_{*}$ shift symmetry. }
  \label {fig:classes}
\end{center}
\end{figure}

The $({\mathbb Z}_5)_*$  discrete shift symmetry admits topological operators such as    
$ e^{-5S_0} e^{i5 \sigma} $. We wish to provide  a physical interpretation of this operator. 
It is apparent that this operator    can only be induced by a  topological excitation with a {\it vanishing} index and with  magnetic charge $+ 5$.
It has the same quantum numbers as a five-monopole state with   three  ${\rm BPS}$ and two   
${\overline {\rm KK}}$ monopoles.     Since each constituent monopole has magnetic charge $+1$, naively,  such an excitation should not be stable, as there is a  pair-wise Coulomb repulsion  between the constituents.  However, there are also interactions induced by the fermion zero-mode exchange, as in the  stability of magnetic bions \cite{Unsal:2007jx}. (A magnetic bion of a newer variety will be discussed in Section 3.) 
This is by no means a simple interaction as it must, at leading order,
be a five-body interaction which glues these constituents into what we call a ``magnetic quintet".  
Schematically, consider the product operator:
\begin{equation}
{\cal Q}  =  [{\cal M}_{\rm 1}  ]^3    [\overline {\cal M}_{2 }]^2   ~,
\end{equation}
and contract  twelve fermion zero modes $\psi^{12}$ in $ [{\cal M}_{\rm 1}  ]^3 $ with the 
twelve opposite chirality fermions   $\bar \psi^{12}$ in  $ [\overline {\cal M}_{2 }]^2$.  We expect  the fermion zero mode exchange to generate a binding potential  (which must be short-ranged, as the fermion zero modes of $\psi$ have an exponential fall-off) for the constituents.

  The magnetic and topological charges of the magnetic quintet are:
 \begin{equation} 
\left( \int_{S_{
\infty}^2} B,  \;\; \frac{1}{32 \pi^2}\int_{\R^3 \times S^1} \; G^a  \widetilde G^a  \right) =  \pm \left(  5, \; \frac{1}{2}\right) 
 \end{equation}
 where the signs are correlated. Its  net number of the fermionic zero modes is zero.  In the 
 effective theory, it generates the operators $e^{\pm 5 i \sigma}$. 

At the end of this Section, we note that while our arguments above are based purely on an analysis of the  symmetries and allowed  flux operators and 
 do not allow us    to establish the existence of ``magnetic quintets" within a controlled analytic approximation, topological objects of magnetic charge 5
could be searched for on the lattice via smoothing, or ``cooling", of lattice field configurations; see \cite{Berg:1981nw} for early references and,  {\it e.g.},  \cite{Bruckmann:2004ib} for more recent work. This is  an interesting direction to pursue in the future and we will only note here that, in the continuum, the fermion determinant of this chiral gauge theory is real and thus  the biggest obstacle to  lattice studies of general chiral gauge theories is absent. 

\section{ Supersymmetric chiral $\mathbf{SU(2)}$ with $\mathbf{I = {3\over 2}}$ matter }
Next, we consider the  supersymmetric ${\cal{N}}=1$ gauge theory with a single chiral superfield in the $I\:$=$\:3/2$ representation. 
This theory was studied in detail by   Intriligator, Seiberg, and  Shenker [ISS] in \cite{Intriligator:1994rx}, where it was shown that 
if this theory exhibits confinement at the origin of the moduli space,  the theory will dynamically break supersymmetry when a tree level superpotential is added. 
By using  recent techniques developed in the non-supersymmetric context \cite{Shifman:2008cx, Poppitz:2008hr}, we will discuss 
the confinement assumption and argue that the theory does not confine.  If so, 
the dynamical breaking of supersymmetry  does not take place in this theory upon the addition of a tree-level superpotential.

Let $Q_{abc} \equiv Q= q + \sqrt 2 \theta \psi + \theta \theta F $ denote the chiral superfield in the $I\:$=$\:3/2$  representation and $\lambda$---the adjoint gaugino.  The basic gauge singlet chiral operator is  $u= Q^4$.
The instanton vertex is:
 \begin{equation} 
 \label{instanton2}
 I(x)= e^{-S_{\rm inst}} \psi^{10} \lambda^4,
\end{equation} 
and an exact  anomaly-free chiral $U(1)_R$ symmetry holds quantum mechanically, under which:
\begin{eqnarray}
 \label{charges1}
&& [\lambda]=+1,\nonumber \\
&& [Q]= {3\over 5}, \qquad    [\psi]=-{2 \over 5}, 
\\ 
 && [u]={12\over 5},  \qquad  [\psi_u]= [q^3\psi]={7\over 5}~.  \nonumber 
 \end{eqnarray}
 Here,  $\psi_u$ denotes  the fermionic component of $u$. 
 
At the classical level, the theory has a moduli space  of degenerate vacua, a Higgs branch parameterized 
 by $u \neq 0$,  along which $SU(2)$ is completely broken. 
  Classically, there is a singularity at the origin  $u =0$, where massless gauge fluctuations appear. Our interest is the dynamical behavior of this asymptotically free theory at the origin of moduli space.

As argued by ISS, there are two logical possibilities at $u=0$ at the quantum mechanical level. The first is a non-abelian Coulomb phase of strongly interacting quarks and gluons, and the other is a confining phase (without chiral symmetry breaking) where the singularity is smoothed out. 
The fact that confinement with   chiral $U(1)_R$ symmetry breaking is not a possibility follows from supersymmetry.  Thus, in neither  of the  two possibilities does  $U(1)_R$ break.
 To show this,  first note that holomorphy and $U(1)_R$ symmetry restrict  the form of a dynamical superpotential to 
$W= c u^{5/6} \Lambda^{-1/3}$, where $\Lambda$ is the strong scale of the theory.  
However, this potential is  incompatible with the weak coupling regime in the moduli space where 
$u^{1/4} \gg \Lambda$, thus $c=0$. Consequently, 
\begin{equation}
W[u]= 0~,
\label{4dW}
\end{equation} 
and the quantum theory, just like the classical theory, has a moduli space of degenerate vacua. A new derivation of  (\ref{4dW})  will be given below. 
  Since $\langle \lambda \lambda \rangle \sim \frac{\partial W}{\partial \tau} =0$, where $\tau$ is the holomorphic coupling, no fermion bilinear condensate forms and $U(1)_R $  remains unbroken. So far, this is all one can say about the dynamics on $\R^4$. 

  \subsection{ Compactification to    $\mathbf{\R^3 \times S^1}$}

 Instead of $\R^4$, we will study $\R^3 \times S^1$ with a periodic spin connection for fermions. 
 This setup  provides the only known  controllable deformation of the  chiral  gauge theory at hand.  As discussed below,   the chiral theory on the circle possesses   a moduli space of vacua composed of a  Coulomb and a Higgs branch and an intersection point where they meet, at the classical level.   We show that neither branch is lifted quantum mechanically. 
At small  $S^1$, we will be able to demonstrate that no mass gap for long distance gauge fluctuations can appear at  the origin of moduli space  and on the Coulomb branch,  hence the theory does not confine. 
One of the main points of the analysis   is that  the non-perturbative consistency of the theory demands, in the semi-classical regime on the Coulomb branch,  intertwining of the $U(1)_R$  symmetry  with a continuous topological $U(1)_J$ shift symmetry for the dual photon. 
This  prohibits an explicit dual mass term for the 
 gauge fluctuations. As will be discussed below, spontaneous breaking of $U(1)_R$ may in principle induce a mass term for gauge fluctuations, however, this possibility is forbidden by 
 supersymmetry. At large $S^1$, we argue that there is most likely  no phase transition 
 on the way and the theory on   $\R^4$ does not confine at the origin of moduli space.   
 
As in the non-supersymmetric theory, there are two types of ``elementary" topological excitations, which we label by ${\cal{M}}_{1,2}$, in the same manner as in (\ref{monopoles}).
 The number of 
fermionic zero modes for these two topological excitations  is given in  \cite{Poppitz:2008hr}:
\begin{equation}
{\cal I}_{1} =(4 \psi, 2 \lambda), \; \;     {\cal I}_{2} =(6\psi, 2 \lambda), \; \; 
{\cal I}_{\rm inst}= (10 \psi, 4 \lambda)~.
\end{equation}
Thus, the corresponding monopole operators are, similar to (\ref{monopoles}):   
   \begin{eqnarray}
 &&  {\cal M}_{\rm 1} = e^{-S_0} e^{-\phi +i \sigma} \psi^4 \lambda^2, \;\;  \overline {\cal M}_{ 1} = 
   e^{-S_0} e^{-\phi - i \sigma} {\bar \psi}^4 {\bar \lambda}^2,  \qquad \cr \cr 
&&    {\cal M}_{\rm 2} = e^{-S_0} e^{+ \phi - i \sigma} \psi^6  \lambda^2   , \;\;
     \overline {\cal M}_{2 } =  e^{-S_0} e^{ + \phi +i \sigma} {\bar \psi}^6  {\bar \lambda}^2~, \qquad 
 \label{monopole2}
   \end{eqnarray}
where $\phi$ is the scalar associated with the holonomy of Wilson line and $\sigma$ is the dual photon (to keep the similarity with the non-supersymmetric case (\ref{monopoles}), but somewhat at odds with the existing supersymmetric literature, we denote by $\phi$ the  expectation value of the holonomy shifted with respect to the center-symmetric value). Note that, as in 
(\ref{inst-mon}),  the product  $ {\cal M}_{\rm 1}   {\cal M}_{\rm 2}$ is just the  instanton vertex (\ref{instanton2}). The Coulomb branch of the supersymmetric theory is parameterized by the chiral  superfield $Y$, which in the semiclassical domain can be expressed as
$Y \sim  e^{-\phi +i \sigma+ \ldots}$.

Since $U(1)_R$ is a true symmetry of the microscopic theory, it must   be a symmetry of the long distance theory (in the small-$S^1$ regime where the long distance theory can be constructed), as well as of all topological operators. Otherwise this would have implied that it is anomalous. Under $U(1)_R$, 
       \begin{eqnarray}
&& \psi^4 \lambda^2 \rightarrow e^{i  \frac{2  \alpha}{5} } \psi^4 \lambda^2,  
 \qquad   
                          \psi^6 \lambda^2 \rightarrow e^{- i  \frac{2 \alpha}{5} } \psi^6 \lambda^2 ~.  
\qquad
 \label{U1R}
    \end{eqnarray}
Thus, the monopole operators are invariant under the $U(1)_R$ if the dual photon transforms by a continuous shift symmetry $U(1)_J$, as opposed to the discrete shift symmetry  (\ref{photonshift}) in the non-supersymmetric case:
       \begin{eqnarray}
  \sigma \rightarrow \sigma -   \frac{2  }{5}    \alpha, \qquad [Y]=-\frac{2}{5} ~.
 \label{continuous}
    \end{eqnarray}
In this sense,  $U(1)_R$   intertwines with the topological continuous shift symmetry  $U(1)_J$ of the dual photon as: 
\begin{equation}
[U(1)_R]_{*}= U(1)_R - \frac{2}{5}U(1)_J \; .
\label{U1*}
\end{equation}
 In the literature,  this step is often implied  and  $U(1)_R$ and $[U(1)_R]_{*}$ are used interchangeably.  For clarity, we wish to distinguish the two. In particular, the conserved current associated with $[U(1)_R]_{*}$  in the long distance 3d theory is, $K_{\mu} = \overline\lambda \overline \sigma_{\mu}\lambda   - {2 \over 5}   \overline\psi \overline \sigma_{\mu}\psi - {2 \over 5} \partial_{\mu} \sigma$, containing terms both of local and topological nature. The conservation of the current 
 $K_{\mu}$ is the local manifestation of the index theorem.  
Therefore, unlike the non-supersymmetric theory, an explicit   mass term for  the dual photon due to a topological operator of the form  $e^{i n \sigma}$   is forbidden  by  $[U(1)_R]_{*}$.
 A parity-odd 
Chern-Simons mass term does not get generated either \cite{Poppitz:2008hr}. This implies that 
the photon and its supersymmetric partners must remain 
 massless in the small-$S^1$ regime.  
 
 Before discussing the strong-coupling  large-$S^1$ regime, we need to know  whether there is any superpotential being generated on $\R^3 \times S^1$.  To do so, we follow the strategy of  
Sections 6 and  7 of ref.~\cite{Aharony:1997bx}.  
 One of the main points of the analysis there is that 
 if a supersymmetric  gauge theory on $\R^3 \times S^1$ has a nonperturbative 
 superpotential $W_{S^1 \times \R^3} [Y, \Lambda, M, \ldots]$ (where $\Lambda$ is the 4d holomorphic strong scale,  $M$ and ellipsis are mesons  and other relevant chiral composites), the superpotential 
  $W_{\R^4} [\Lambda, M, \ldots]$ (or  quantum moduli space) on 
 $\R^4$ can be obtained by integrating out $Y$, the superfield 
 associated with the Coulomb branch  
 on $S^1 \times \R^3$   (this is because   $\sigma$ parameterizes an $S^1$ and 
 $\phi$   parameterizes   $  S^1 / \Z_2 \equiv I \equiv [0, \pi]$, the $\Z_2$ orbifold of $S^1$ by the Weyl group  $\Z_2$ of $SU(2)$, whose size shrinks to zero in the $\R^4$ limit, see, {\it e.g.},~\cite{Seiberg:1996nz}). This  implies that the vacuum structure of a supersymmetric gauge theory 
 on  $S^1 \times \R^3$ can be used to deduce the vacuum structure of the same theory on $\R^4$, i.e, 
 \begin{equation}
 W_{S^1 \times \R^3} [Y, \Lambda, M, \ldots] \underbrace{\longrightarrow}_{{\rm \; Integrate \; out}\;  Y} W_{\R^4} [ \Lambda, M, \ldots] \qquad {\rm {\it or}\; \;  \;quantum \; moduli \;  space \; constraint}.
 \label{integrate}
 \end{equation}
This assertion is true for all supersymmetric theories studied in \cite{Aharony:1997bx}, and we believe it holds in general.  As a side note, we wish to point that an analog of this statement, smoothness of physics as a function of radius is also achieved for certain non-supersymmetric vector-like and chiral gauge theories by using double-trace deformations 
\cite{Shifman:2008cx}. Of course, the beauty in both  supersymmetric and non-supersymmetric cases is that we can connect a strongly coupled dynamical regime into a semi-classically 
tractable regime, where we can essentially solve the theory.  This is the importance of studying gauge theories on ${S^1 \times \R^3}$.

The way to obtain  $W_{S^1 \times \R^3} [Y, \Lambda, M, \ldots]$  
 is to start with the theory on $\R^3$, find the superpotential on $\R^3$ and add to it 
 any contribution that may arise due to extra topological excitations  inherent to compactification \cite{Aharony:1997bx}. 
  
 \subsection{ Supersymmetric $\mathbf{I = {3\over 2}}$ theory on  $\mathbf{\R^3}$}
 
  Since chiral anomalies are not present in odd dimensions, the $U(1)_R$ symmetry of the locally four dimensional theory enhances to  $U(1)_{R'} \times U(1)_A$ upon dimensional reduction to  $\R^3$. 
  The superpotential of the three-dimensional theory is constrained by the global symmetries 
(\ref{U1*two}), under which the charges are as follows:
 \begin{equation}
\begin{array}{ccc}
& [U(1)_{R'}]_{*} & [U(1)_A]_{*} \cr
\lambda & 1 & 0 \cr
\psi & -1 & 1\cr
Q & 0& 1\cr 
Y & 2 & -4 
\end{array}~.
\label{3dsym}
 \end{equation}
 The charges of $Y$ under the global symmetries can be inferred in many ways, as explained in \cite{Aharony:1997bx}, one of which is to evoke  the  index theorem in monopole backgrounds on $\R^3$ \cite{Poppitz:2008hr}. This is because the invariance of  ${\cal M}_1 \sim e^{- \phi + i \sigma} \lambda^2 \psi^4$  implies,  as  in (\ref{u1j}), that these local symmetries intertwine with the topological symmetry as:
   \begin{equation}
[U(1)_{R'}]_{*}= U(1)_{R'} + 2 U(1)_J , \qquad [U(1)_A]_{*}= U(1)_A - 4 U(1)_J~, 
\label{U1*two}
\end{equation}
thus determining the charge of $Y$ in (\ref{3dsym}).  However, 
notice that   ${\cal M}_2$ ($\overline {\cal M}_2$) is not invariant under   (\ref{3dsym}) 
 as the KK monopoles  do not exist in the gauge theory on $\R^3$.

  In terms of the superfields  $u = Q^4$ and $Y$,  parameterizing the Higgs and Coulomb branches ${\mathbb C} \times (\R^{+}\times S^1)$, 
there is a unique superpotential permitted   by symmetries and holomorphy: 
       \begin{eqnarray}
 W[Y, u] =  b \;Yu~.
 \label{superpotential}
    \end{eqnarray}
 This type of superpotential is reminiscent of the ones studied   in the context of SQCD in 
 Section 6 of \cite{Aharony:1997bx}.  
 The superpotential (\ref{superpotential}) as well as the 
 ones in various vectorlike theories from ref.~\cite{Aharony:1997bx}  are permitted by symmetries,  but their   origin   is not  yet discussed in the literature. 
Here, we would like to discuss  the microscopic origin of this superpotential as well as point out the difference between the 
vectorlike and chiral cases.

\begin{figure}[ht]
\begin{center}
\includegraphics[angle=0, width=0.7\textwidth]{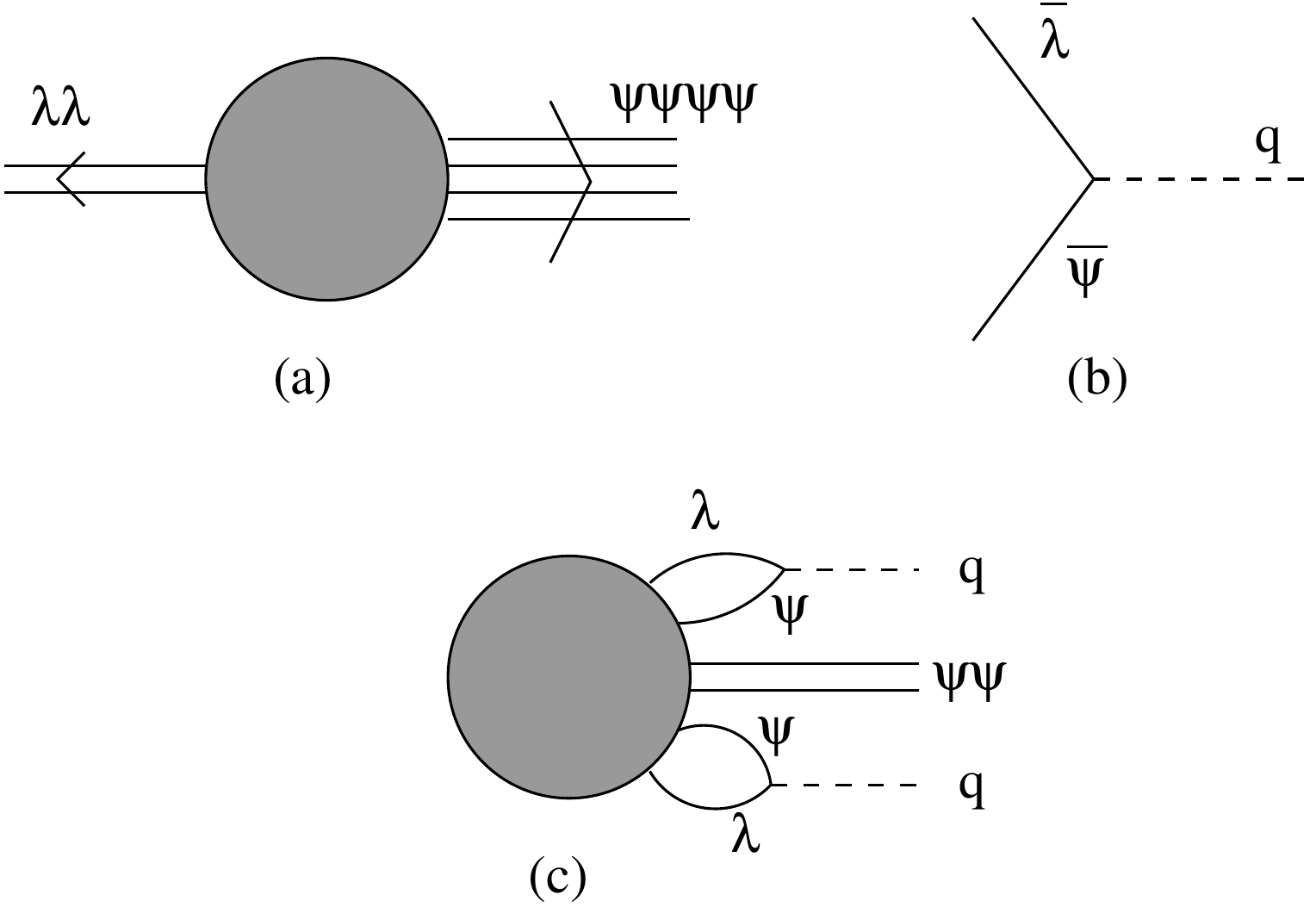}
\caption{
(a) is the  monopole operator ${\cal M}_1$, e.g., (\ref{monopole2}), (\ref{monQCD}) dictated by the index theorem. 
(b) is the  Yukawa vertex. (c) is a modified monopole operator   $ \widetilde 
 {\cal M}_{\rm 1}$  e.g, (\ref{monopole2mod}), (\ref{monQCD2}) obtained upon Yukawa contractions. Note that  $ \widetilde 
 {\cal M}_{\rm 1}$  has exactly  two fermionic zero modes and  can thus contribute to the superpotential.     
 }
  \label {fig:monop}
\end{center}
\end{figure}

 Recall the monopole operator (\ref{monopole2}) $  {\cal M}_{\rm 1} = e^{-S_0} e^{-\phi +i \sigma} \psi^4 \lambda^2$ pertinent to the Coulomb branch of the gauge theory on $\R^3$. The structure of the fermion zero modes is dictated by the index theorem and, as it stands,  $  {\cal M}_{\rm 1}$ has more than two   zero modes and cannot  contribute to a superpotential. 
However,  this argument does not take into account the Yukawa interaction, which does not enter into the index theorem and  which may lift zero modes. 
The Yukawa interaction is of the form 
$ q  \bar \lambda \bar \psi  + {\rm h.c.} $.   Contracting the Yukawa interaction (twice) with the monopole operator soaks up  
two $\lambda$ and two $\psi$ zero modes and introduces two scalars, as shown in 
Fig.~\ref{fig:monop}:
\begin{eqnarray}
 e^{-S_0} e^{-\phi +i \sigma} \psi^4 \lambda^2 (x) \left( \int d^3y \;  q  \bar \lambda \bar \psi  (y) \right)^2  \longrightarrow    
 \widetilde 
 {\cal M}_{\rm 1}  \equiv  e^{-S_0} e^{-\phi +i \sigma} q^2 \psi^2  ~.
 \label{monopole2mod}
 \end{eqnarray}
A possible non-locality of the integrals is cut-off by the Coulomb-branch mass term in  the $\psi$ propagator at large distances, while, at short distances, the monopole size puts a natural cut-off. 
  The resulting 
expression can be viewed as a  {\it modified monopole operator}   $\widetilde 
 {\cal M}_{\rm 1} $,  with just two zero modes, an exemplar of non-perturbatively generated  superpotential (\ref{superpotential}):
      \begin{eqnarray}
 W[Y, Q] \sim  Y Q^4, \qquad  \widetilde 
 {\cal M}_{\rm 1}   = \frac{\partial^2 W}{\partial q^2}  \psi \psi~,
 \label{superpotential2}
    \end{eqnarray} 
in the semi-classical domain.   
There are well-known textbook examples   where instantons  on  $\R^4$  produce a non-perturbative superpotential on the   Higgs branch of SQCD,  see for example \cite{Shifman:1999mv}.  The difference  here is that the above modification of the monopole operator takes place on the Coulomb branch, {\it i.e.}, with a real representation  (adjoint) Higgs vev insertion,  
resulting in   reduced (local) monopole operators  from the viewpoint of the long-distance theory. 

At this stage, it is also worth noting that   in non-supersymmetric center-stabilized QCD-like and chiral gauge theories, there are no Yukawa interactions. Hence, the zero modes of the leading monopole operators  do not get lifted on $\R^3$ and $\R^3 \times S^1$, as in supersymmetric theories.  This makes  the analysis of the deformed YM theories with fermionic matter slightly more  friendly than the supersymmetric theories 
\cite{Shifman:2008cx,Unsal:2007jx}.

Returning to  superpotential (\ref{superpotential2}),  it leads  to the $F$-term  bosonic potential, given schematically by:
   \begin{eqnarray}
 V _F(\phi, q) \sim   e^{-2S_0} e^{-2\phi}  q^6 (1 + {\cal O}(q^2)  )
  \label{pot1}
    \end{eqnarray} 
Note that (\ref{pot1}) is independent of $\sigma$, as it must, because the 3d theory possesses two  shift symmetries  (\ref{U1*two}) under which   $\sigma   \rightarrow \sigma  + 2 \alpha_{R'}$ and    $\sigma   \rightarrow \sigma  - 4 \alpha_{A}$, as is manifest by the charge assignments shown in (\ref{3dsym}). It is clear that the Coulomb branch is not lifted by the potential (\ref{pot1}), which vanishes for 
  $q=0$ and arbitrary $\phi >  0$. The Coulomb branch is expected to persist in the strong coupling domain as well. 
  
An important question that must be asked before deciding whether (\ref{superpotential}) is relevant for describing the long-distance dynamics of the theory anywhere on moduli space is
whether there is a region of the moduli space where both $Y$ and $u$ comprise the light degrees of freedom. If so, one would conclude that there
 (\ref{superpotential}) is the correct Wilsonian superpotential of the theory and, by holomorphy, extend it over the entire moduli space. It is clear that  on the Coulomb branch $u$ is heavy, as adjoint  scalar expectation values  give large mass to all components of $Q$,  and that, conversely,  on the Higgs branch the entire gauge multiplet, and thus  $Y$, is heavy. Hence, we are forced to study the origin of moduli space and ask  whether the long distance regime of the theory on $\R^3$ may be described   in terms of the $Y$ and $u$ fields alone (in other terms, if the 3d theory is confining, with the original gluons, quarks, and superpartners not appearing in the long-distance description).  If so, the microscopic 
 discrete parity anomalies must match to the macroscopic ones.  Below, we demonstrate a mismatch. The parity anomaly is defined as: 
 \begin{equation}
 k_{ij}= \half \tr(q_iq_j)= \half \sum_{f} q_{f,i}q_{f,j}~,
 \end{equation}
 where $q_{f,i}$ is the charge of the fermion $f$ under $U(1)_i$ and the sum is over over all fermions. For microscopic anomalies, we find:
\begin{eqnarray}
&&k_{R'R'} = \frac{1}{2}\left[3  (1)^2+ 4  (-1)^2\right] =  \frac{7}{2} \in {\Z + \half} ~,  \nonumber \\
&&k_{R'A} = \frac{1}{2}\left[3 (1 )( 0)+ 4  (-1)(1)\right] =  -2  \in {\Z} ~,  \nonumber \\
&&k_{AA} = \frac{1}{2}\left[3 (0)^2+ 4  (1)^2\right] =  2 \in {\Z} ~,
\end{eqnarray}
by adding the contributions of the   three adjoint fermions $\lambda$ and the four components of the
$I=3/2$ fermions $\psi$.  
The macroscopic anomalies of the fermionic components   $(\psi_Y, \psi_u)$
of the $(Y, u)$ superfields  are:
\begin{eqnarray}
&&k_{R'R'} = \frac{1}{2}\left[1 (1)^2+ 1 (-1)^2\right] =  1 \in {\Z} ~,  \nonumber \\
&&k_{R'A} = \frac{1}{2}\left[1 (-4 )( 1)+ 1 (-1)(4)\right] =  -4  \in {\Z} ~,  \nonumber \\
&&k_{AA} = \frac{1}{2}\left[1 (-4)^2+ 1  (4)^2\right] =  16 \in {\Z}~.
\end{eqnarray}
Due to the mismatch of the  $k_{R'R'}$ anomalies, the $(Y,u)$ fields cannot provide 
a consistent description of the long distance theory near the origin of moduli space. Other degrees of 
freedom are required in order to match the parity anomalies and the theory at the origin is most likely 
a strongly coupled CFT of the original $I=3/2$ ``(s)quarks" $Q$ and the $SU(2)$ (s)gluons and gluinos. 
Thus, there is no known region of moduli space where both $Y$ and $u$ are the light degrees of freedom and we conclude that in (\ref{superpotential}) $b=0$ (note also that  this is a somewhat foregone conclusion as (\ref{superpotential}) with $b \ne 0$ would then  be a mass term).

The ``chiral" $I=3/2$ theory on $\R^3$ in this sense differs from the vector-like SQCD examples studied in  \cite{Aharony:1997bx}.  
The simplest example studied  there, which is useful for comparison with the $I=3/2$ chiral case of interest,  is  $N_f=2$ $SU(2)$ SQCD on    $\R^3 $ and  $\R^3 \times S^1$; for brevity, we will often call the latter theory ``the vectorlike theory."
 
Let us first  
 discuss  the microscopic origin of  the superpotentials in the vectorlike theory on $\R^3$.
  In this case, there is also a unique superpotential permitted by symmetries and consistent with holomorphy. It is expressed in terms of the $Y$ chiral superfield labeling the  Coulomb branch 
 and the meson chiral superfields, $M_{ab}= Q_a \cdot Q_b$, ($a, b=1, \ldots,  4$) labeling the  Higgs branch, and is given by: 
\begin{equation}
W= -Y\;  {\rm Pf }(M) = -Y M_{12} M_{34} + \cdots~,
\label{SQCD}
\end{equation}
where ${\rm Pf }$ is the  Pfaffian. 
A microscopic derivation of   (\ref{SQCD}) can be given  following the same line of reasoning as in the chiral $I=3/2$ theory. 
The  monopole operator   on  the Coulomb branch of the vectorlike theory is: 
 \begin{equation}
   {\cal M}_{\rm 1} = e^{-S_0} e^{-\phi +i \sigma} \psi_1\psi_2 \psi_3 \psi_4 \lambda^2 \; .
   \label{monQCD}
   \end{equation}
As usual, the structure of zero modes is dictated by the index theorem. Exactly as in the chiral $I=3/2$ theory, there are too many zero modes, but the theory also has Yukawa interactions, $ q_a  \bar \lambda \bar \psi_a  + {\rm h.c.} $, 
which  lift fermion zero modes in pairs and introduce scalars for each pair, see 
Fig.~\ref{fig:monop}, resulting in the modified monopole operator:
\begin{eqnarray}
 \widetilde  {\cal M}_{\rm 1}  \equiv
 e^{-S_0} e^{-\phi +i \sigma} ( q_1q_2 \psi_3\psi_4  +  \ldots) =  \sum_{a, b} \frac{\partial^2 W}{\partial q_a\partial q_b}  \psi_a \psi_b \; .
     \label{monQCD2}
\end{eqnarray}
Here, the ellipsis stands for other permutations and $W$ is the superpotential given in  (\ref{SQCD}). 
This expression is valid in the  semi-classical domain along the Coulomb branch.  The bosonic potential is formally similar to (\ref{pot1}), hence 
neither   the Coulomb, nor the  Higgs branch is  lifted in the vectorlike theory 
 on $\R^3$. The difference with the chiral case is that, at the origin, the fields $Y$ and $M_{ab}$ saturate the parity anomaly matching condition---hence it was argued in  \cite{Aharony:1997bx}  that at the origin of moduli space the theory is dual to  a chiral superfield CFT with superpotential (\ref{SQCD}), {\it i.e.}, to a supersymmetric 3d Wilson-Fisher fixed-point theory. By holomorphy of the Wilsonian effective action the superpotential extends to the entire field space of the low-energy theory.
 
\subsection{Back to    $\mathbf{\R^3 \times S^1}$}

  We now proceed to study the chiral and vectorlike theories of the previous section on $\R^3 \times S^1$.  
For  the vectorlike supersymmetric  theories studied in \cite{Aharony:1997bx}, compactification on 
 $S^1$ always induces a term in the superpotential linear in $Y$, $\delta W = \eta Y$. This is due to the fact that all fundamental matter zero modes are localized into one topological excitation    (a BPS magnetic monopole, the analog of 
 ${\cal M}_1$), while the other topological excitation  (the KK monopole, ${\cal M}_2$) only carries two adjoint zero modes, and thus generates an operator ${\cal M}_2 \sim e^{+ \phi - i \sigma}  \lambda^2$. Consequently, it   contributes an   $\eta Y$ deformation to the three dimensional superpotential.

 In the $N_f=2$ $SU(2)$ vectorlike theory described above, these two type of monopole operators induce terms like:
 \begin{eqnarray}
 \widetilde  {\cal M}_{\rm 1} + {\cal M}_{\rm 2} = 
 e^{-S_0} e^{-\phi +i \sigma} ( q_1q_2 \psi_3\psi_4  +  \ldots)  +  e^{+ \phi - i \sigma}  \lambda^2~,
 \end{eqnarray}
 which naturally arise from the superpotential  proposed in ref.~\cite{Aharony:1997bx}:
 \begin{equation}
W= -Y\;  {\rm Pf }(M)  + \eta Y~.
\label{SQCD2}
\end{equation}
Integrating out $Y$  from (\ref{SQCD2}), as schematically shown in Eqn.~(\ref{integrate}), then gives rise to a quantum modified moduli space   ${\rm Pf }(M)  = \eta$,
 the correct result on  $\R^4$ \cite{Seiberg:1994bz}.
 The $F$-term bosonic potential on  what used to be the Coulomb branch of the $\R^3$-theory is modified on $\R^3 \times S^1$ 
 into (schematically):
    \begin{eqnarray}
 V_F(\phi, q) \sim   e^{-2S_0} e^{-2\phi}  q^6 (1 + {\cal O}(q^2)  ) +  e^{-2S_0} e^{-2\phi} ~,
  \label{pot2}
    \end{eqnarray} 
 and the KK-monopole induced superpotential appears to generate a ``run-away" potential for the $\phi$ field. However, since the $\phi$-space 
 is compact for the gauge theory on $\R^3 \times S^1$, this just means that the two eigenvalues of the Wilson line (located at $\pm \phi$) will merge together. This regime is highly quantum, meaning that it does not admit a semi-classical description.   Even staying  within the semi-classical domain and setting  $q=0$, we see that the potential  
$  V_F (\phi, 0) \sim  e^{-2\phi} $ is non-vanishing, the  Coulomb branch is lifted, and the vacua are located  at $Y=0$. The field $Y$ thus obtains mass while the fields $M_{ab}$ remain massless, subject to the quantum-modified constraint. Thus, the $L=0$ and $L >0$ theories differ in the sense that $Y$ is massive in the latter. In the vectorlike theory, the  
 absence of massless non-abelian gauge  fields ({\it i.e.},~non-abelian confinement) 
persists  on $\R^3 \times S^1$ at any $L>0 $ and the dynamics smoothly connects to the known 4d result.

 In the chiral case, the situation is quite different, as we already saw on $\R^3$. Most importantly, the Coulomb branch on $\R^3$ is not lifted by nonperturbative effects on $\R^3 \times S^1$. This is because the extra topological excitation ${\cal M}_2$, shown in 
 (\ref{monopole2}) for our theory, has six matter and two adjoint zero  modes, as opposed to just two adjoint zero modes in SQCD.  Yukawa interactions will again lift matter and adjoint zero modes in pairs, leading to a modified monopole operator    $\widetilde {\cal M}_2$.    For comparison with 
 (\ref{monopole2}), we collect below the formulae for the modified monopole operators in the $I=3/2$ theory: 
     \begin{eqnarray}
 &&  \widetilde {\cal M}_{\rm 1} = e^{-S_0} e^{-\phi +i \sigma} q^2 \psi^2, \;\;\;   \widetilde{\overline {\cal M}_{ 1}} = 
   e^{-S_0} e^{-\phi - i \sigma} {\bar \psi}^2 {\bar q}^2,  \qquad \cr \cr 
&&     \widetilde  {\cal M}_{\rm 2} = e^{-S_0} e^{+ \phi - i \sigma} \psi^4  q^2   , \;\;\; 
      \widetilde {\overline {\cal M}_{2 }} =  e^{-S_0} e^{ + \phi +i \sigma} {\bar \psi}^4 {\bar q}^2. \qquad 
 \label{modmon}
   \end{eqnarray}
   As stated above, the KK-monopole induced 
$ \widetilde {\cal M}_2 $ has  too many zero modes to contribute to the superpotential,  
  hence it does not.  Thus, the vanishing superpotential 
 on $\R^3$ is  not  modified on $\R^3 \times S^1$ (of course,  the locally four dimensional theory  only has an 
   $U(1)_R = U(1)_{R'} + \frac{3}{5} U(1)_A$ linear combination of symmetries due to the anomaly, 
   and  (\ref{superpotential}) is automatically invariant under  $[U(1)_R]_{*}$). This implies that the 
  Coulomb branch persists quantum mechanically, unlike the case of SQCD described above.  In the decompactification limit, we may integrate out  $Y$, and since there is no superpotential on $\R^3 \times S^1$,  we  obtain $W_{\R^4}[u]=0$---giving a new derivation of the result (\ref{4dW}) of \cite{Intriligator:1994rx}.

\begin{figure}[ht]
\begin{center}
\includegraphics[angle=0, width=0.7\textwidth]{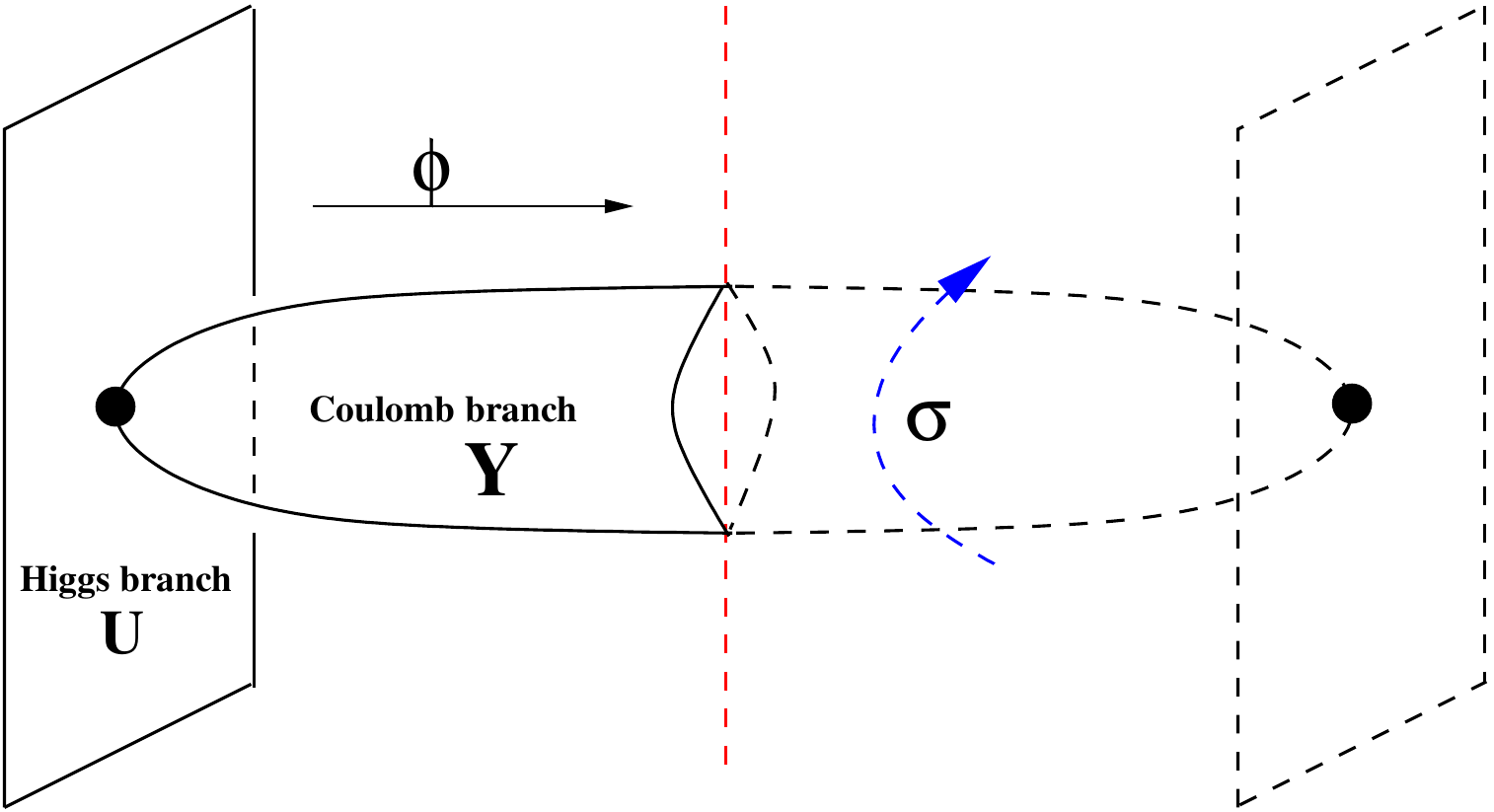}
\caption{The moduli space of the chiral supersymmetric $SU(2)$ $I=3/2$ theory on $\R^3 \times S^1$, with Higgs branch parameterized by $u \in \mathbb C$, and a Coulomb branch  parameterized by $Y$. In the semi-classical domain, $Y \sim e^{-\phi +i \sigma}$, where $(\phi, \sigma) \in (S^1/\Z_2) \times S^1$. The geometrized   $\sigma$ rotation is the topological $U(1)_J$ symmetry.  The fixed point of the $\Z_2$ action shown by red-dotted line is the center-symmetric configuration. On $\R^3$,  
 $(S^1/\Z_2) $ is replaced by $\R^{+}$, as the $\phi$ direction decompactifies into a semi-infinite cylinder. In the $\R^4$ limit, the Coulomb branch shrinks to zero and only the Higgs branch survives. 
 }
  \label {fig:moduli}
\end{center}
\end{figure}

We conclude that the moduli space  of the chiral theory on $\R^3 \times S^1$ is not lifted, thus no fields massless in 3d  acquire mass upon ``turning on" a small-radius $S^1$---as $Y$ does in the vectorlike case, recall  (\ref{SQCD2}) and discussion thereafter---and hence the long-distance physics of the theory is unaffected. Since, 
as we argued above, on $\R^3$  the theory flows to a fixed point at the origin of moduli space and a mass gap  in the gauge sector is   absent on the entire Coulomb branch, we  expect the absence of mass gap and confinement to hold also at sufficiently small radius of $S^1$, over the entire Coulomb branch.

 {\flushleft \it Comments on the semi-classical domain of Coulomb branch:} 
 
The form of the monopole operators (\ref{monopole2}) may also suggest that the Coulomb branch
could be lifted by spontaneous breaking of $U(1)_R$ symmetry, and this may generate a mass for the dual photon and hence cause  confinement.   There are 
 two crucial  observations which we will use in exhibiting the absence of such confinement mechanism in this theory:
 \begin{itemize}
 \item{ Due to the index theorem, $[U(1)_R]_{*}$ acts as a  topological  continuous shift symmetry (\ref{continuous}), forbidding any explicit mass term   for the  dual photon.  
 This is a necessary, but not a sufficient condition.}  
 \item{ 
 A mass term for the dual photon could be generated by dynamical breaking of $U(1)_R$, as indicated below. However, in the supersymmetric $I=3/2$ theory, spontaneous $R$-symmetry breaking is forbidden by supersymmetry.}
 \end{itemize}
 To elucidate the second point above, consider for example the monopole-generated multi-fermion interaction (\ref{monopole2}).  This  interaction becomes stronger as the size of $S^1$ is increased, and depending on the details of the theory, such interactions  could lead to dynamical $R$-symmetry breaking:
 \begin{equation}
  e^{- \frac{8 \pi^2}{g^2(L)N}} e^{i \sigma} \langle \psi^4 \rangle \langle  \lambda^2 \rangle + {\rm c.c} \; \sim \;  e^{-S_0} \cos {\sigma}\; ,  
  \label{cde}
 \end{equation}
generating   a dual  mass term for the   ``the photon component" of the $SU(2)$ gauge fluctuations.  
  However, as argued on  $\R^4$ in  \cite{Intriligator:1994rx}, in the supersymmetric theory, the spontaneous breaking of    $U(1)_R$ is not compatible with holomorphy and weak coupling limits, hence     $\langle \lambda \lambda \rangle  =0$.  This argument can be extended to $\R^3 \times S^1$, by using the fact that  the fermion bilinear  $\lambda \lambda$ is an element of  the chiral ring---the class of operators annihilated by a supercharge  $\overline Q_{\dot \alpha}$ of one chirality---and as such its value is independent of the size of the $S^1$ circle \cite{Witten:2003ye}. 
  Consequently, the gaugino condensate must vanish at any value of radius (furthermore, in our example (\ref{cde}) an expectation value of $\langle \psi^4\rangle$ is forbidden by supersymmetry). 
  Since the $U(1)_R$ symmetry and the topological $U(1)_J$ shift symmetry of the dual photon are intertwined in the semi-classical domain of the Coulomb branch, and since $U(1)_R$ is unbroken, 
   this implies that  ``the photon component" of the $SU(2)$ gauge fluctuations cannot acquire a non-perturbative mass via spontaneous breaking of $U(1)_R$ at any radius where the semi-classical  
   approximation is valid. 
       If the theory has a weakly coupled infrared fixed point, then this radius can be taken to arbitrarily large values, and the intertwining of 
   $U(1)_R$-$U(1)_J$ symmetries is valid at any finite radius. Otherwise, if the theory has a strong scale or flows into a strongly coupled fixed point in the IR,   the notion of dual photon and topological symmetry are useful only in the  $L \ll \Lambda^{-1}$ semi-classical domain, where $SU(2)$  reduces  to $U(1)$.
   This just means that the local 4d dynamics  is unable, under the above conditions, to produce a mass gap for gauge fluctuations, which is in essence protected by an unbroken mixture of 
   $U(1)_R$ and $U(1)_J$ topological symmetry. 
   
   One may also wonder, if there is any gauge theory, which satisfies the first condition, but fails the second, hence, spontaneous breaking of some chiral  $U(1)$ symmetry (analog of $U(1)_R$) may cause mass gap for gauge fluctuations. Indeed, in non-supersymmetric theories, there are such examples. We will report on this interesting class of gauge theories in the future.

 \subsection{ Decompactification to     $\mathbf{\R^4}$}
 
In studying the approach to $\R^4$,  the order of two scales, implicit in our discussion so far, needs to be interchanged: the Kaluza-Klein scale, $M_{KK} \sim {1\over L}$,  and the four-dimensional strong-coupling scale, $\Lambda$, the scale where the 4d theory would become strong, or the scale where the 4d coupling approaches a fixed-point value, which may or may not be small if this theory is conformal, see discussion below. 
The considerations of the previous Section are valid in the $\Lambda \ll M_{KK}$ limit, such that the 4d  gauge coupling  is weak at the scale of the compactification. We argued that, in this limit, the long-distance theory is a supersymmetric CFT of the original fields  of the $SU(2)$ $I=3/2$ theory. 
The main issue in taking the decompactification limit is whether the infrared (below the lowest of  $M_{KK}, \Lambda$) dynamics changes drastically as  one transitions from the $\Lambda \ll M_{KK}$  to the  $M_{KK} \ll \Lambda$ regime, as, clearly, the latter regime is the one relevant to  the 4d limit of interest. 

If the confining hypothesis in 4d held true, such that at scales below $\Lambda$ the 4d theory was that of a free field $u$, with irrelevant K\" ahler potential corrections  $\sim {(u^\dagger u)^2 \over \Lambda^2}$, an $\R^3 \times S^1$
compactification with $M_{KK} \ll \Lambda$ would result in a three dimensional  theory of a single free chiral superfield $u$, 
with irrelevant K\" ahler potential corrections $\sim {(u^\dagger u)^2 M_{KK} \over \Lambda^2}$ (after  3d normalization) and an accompanying Kaluza-Klein tower. Thus, in the confining scenario, in the $M_{KK} \ll \Lambda$ limit one  expects the infrared behavior of a free 3d chiral-superfield theory. 
This behavior is quite distinct from  the one argued for in the $\Lambda \ll M_{KK}$ limit---that of a strongly interacting CFT of the $I=3/2$ (s)quarks, and the $SU(2)$ (s)gluons and gluinos.  Thus one expects that a drastic, possibly discontinuous, change of various long-distance correlation functions should occur as $\Lambda$ crosses the $M_{KK}$ threshold.
 
In contrast, if the 4d $I=3/2$ $SU(2)$ theory was at a fixed point at some scale below $\Lambda$, compactification with $M_{KK} \ll \Lambda$ would lead,   below $M_{KK}$, to an $I=3/2$ $SU(2)$ theory at a fixed point and one would expect that this 3d CFT is continuously---in the sense that there wouldn't be any discontinuous changes in long-distance correlation functions---connected (or even equivalent to) to the 3d $I=3/2$ $SU(2)$ CFT  which we argued to occur when $\Lambda \ll M_{KK}$.

 Admittedly, we have no solid proof that it is the latter option which is chosen. In favor of the second option, we note that the index of the $I=3/2$ representation is rather high. Thus,  see  \cite{Shifman:1999mv}, by comparing the one- and two-loop contributions to the beta function, we can infer a fixed point occurring at a relatively small coupling, ${g_*^2 \over  4 \pi^2}=  {4\over 75} \simeq 0.05$. It is easy to check that the known three- and four-loop corrections to the beta function in the NSVZ-scheme  \cite{Jack:1996cn} do not significantly change this conclusion (we also note that, in contrast, in the non-supersymmetric case this fixed-point value would be the much larger ${32\over 45} \simeq 0.7$). 
 Thus, it is quite likely that the 4d theory flows to a (not too strong) CFT. Upon compactification, this is consistent with our 3d CFT picture.  Also note that  the  diluteness of the  monopole operators (say at the center symmetric point) is controlled by  $e^{-S_0}  = e^{ - \frac{4\pi^2}{g^2(L)}} \leq e^{ - \frac{4\pi^2}{g_*^2}}$, hence for the supersymmetric theory the maximum value that  $e^{-S_0}$  may acquire is $e^{- {75\over4}}$, whereas for the non-supersymmetric theory it is  $e^{- {45\over32}}$. This means that the magnetic monopoles  (and quintets) in the non-supersymmetric theory will  indeed become non-dilute (more relevant) with increasing radius, whereas for the supersymmetric theory, the monopoles always remain arbitrarily  dilute and hence irrelevant. They will eventually be washed-out in the renormalization group sense at long distances.

 More non-rigorous, however more widely used,   arguments against the confining option can be put forward by recalling the discussion two paragraphs above---that in the confining scenario the infrared physics (below min($M_{KK}, \Lambda)$) has a discontinuous behavior as the ratio of two
  scales, $M_{KK}$ and $\Lambda$, is changed. In non supersymmetric theories such discontinuous behavior is common. In supersymmetric theories there is some lore \cite{Seiberg:1994aj, Intriligator:1994sm}  about the absence of phase transitions, based on holomorphy and the ensuing fact that singularities of the superpotential and the holomorphic gauge coupling are   of codimension two  and therefore one can always ``go around" them.  Although the smoothness of physics in supersymmetry preserving compactification of  supersymmetric  gauge theories on  $\R^d \times S^1$ as a function of   radius (or equivalently, as a function of the holomorphic coupling  $\tau(L)$)  is not a theorem,  there are currently no known counter-examples.  While it is not entirely clear to us if this should apply to the present case, we note that   similar cases have been studied in the literature and  smooth behavior  of the infrared physics upon changing the ratio of parameters  has always been found in supersymmetric theories. In this respect, we note the studies \cite{Poppitz:1996vh, Barnes:2005zn} of renormalization group flows in
   product-group theories, where flows that would imply a separatrix in the space of ultraviolet couplings were shown to not occur. Also, quite similar to our present case, the study  \cite{Aganagic:2001uw}  of $S^1$ compactifications of 3d supersymmetric theories argued for the absence of a phase transition as the compactification scale was varied from $g_3^2   \ll M_{KK}$ to $M_{KK} \ll g_3^2$. Studies of  SQCD \cite{Aharony:1997bx} on $\R^3 \times S^1$  and $\R^4$ also provide supporting evidence for smoothness conjecture. 
Smoothness of supersymmetric gauge dynamics as a function of non-vanishing compactification radius suggests  that the disfavored possibility in ref.~\cite{Intriligator:1994rx}, 
   i.e., an interacting strongly coupled CFT at the origin of the moduli space is, in our opinion,   the    expected dynamical behavior.     Indeed, a more recent diagnostic by Intriligator 
    \cite{Intriligator:2005if}, based on $a$-maximization, also suggests that the infrared theory at $u=0$ should be  a strongly coupled CFT. An infrared  CFT  behavior was also advocated earlier in  \cite{Shifman:1999mv}, by arguing that the  $I=3/2$ theory  may lie in the conformal window. 
 
 One may ask why  confinement is argued as the more likely dynamical behavior in 
 \cite{Intriligator:1994rx}.  This assertion is based on 't Hooft anomaly matching of the 
microscopic  and macroscopic anomalies, 
\begin{eqnarray}
&&\tr R= 3 \times 1+ 4 \times \left(-{2\over 5}\right)= {7\over 5} ~,  \nonumber \\
&&\tr [R^3]= 3 \times1^3+ 4\times \left(-{2\over 5}\right)^3= \left({7\over 5}\right)^3~,
\end{eqnarray}
where the l.h.s. of the two equations receives a contribution from the  three adjoint fermions $\lambda$ and the four components of the
$I=3/2$ fermions $\psi$, while the r.h.s. is saturated by the single fermionic $\psi_u$ component of the composite gauge singlet $u$, with  $[\psi_u]=7/5$, as shown in  (\ref{charges1}). 
In this sense, this theory also provides an example 
of  misleading anomaly matching, see \cite{Brodie:1998vv}.   Of course, 
if  the theory at the origin of moduli is a CFT, then microscopic 
and macroscopic fermions are the same and 
 the  't Hooft anomalies  are trivially satisfied.

If the infrared theory is a CFT at $u=0$, then the tree level potential  $\Delta W[u] = \xi u$ is an irrelevant deformation  
 and does not alter the infrared physics. This means that no dynamical supersymmetry breaking takes place.  
 It would be useful to re-examine other models of dynamical supersymmetry breaking which rely on the confinement assumption by using the techniques of this work.
      
\subsection{Comments of  conformal dynamics in 3d and a new class of CFTs} 

On small $\R^3 \times S^1$ and on $\R^3$, we have provided strong evidence suggesting that 
the $I=3/2$ $SU(2)$ supersymmetric theory at the origin of moduli space is similar to the classical theory, in the sense that  massless quarks and gluons are not confined. Evidence gathered in the previous section also suggests that the same conclusion is also true on $\R^4$.  Below, we wish to make various remarks regarding such conformal field theories on $\R^3$ and small  $\R^3 \times S^1$. Our discussion will apply rather generally (without specializing to supersymmetry), and we will point out a novel class of CFTs in 3d. 

Consider YM theory in 3d  with fermions in arbitrary representation  $R$. The  classical 3d theory, as it is superrenormalizable,  possesses a dimensionful coupling $g_3^2$. If such a theory flows into a CFT, the  CFT 
description can arise at low energies or large $g_3^2$, ${E \over g_3^2} \ll1$. In this limit, the 
dimensionful parameter must disappear from the dynamics, due to screening effects of matter 
fields. It is known that for large number of fermionic fields, such theories will flow to a CFT 
\cite{Appelquist:1989tc}. Here, we propose an alternative---motivated by our study of $I=3/2$  representation fermions---where we can reach a perturbatively accessible CFT by taking advantage of higher representations matter fields, while keeping  $N_f=1$ or few. 

It is well-known that 
integrating out a slice of high-momentum modes of fermions (at one loop order) 
alters the gauge kinetic term as:
\begin{equation}
\frac{1}{4g_3^2} \tr F_{\mu \nu}^2 + \ldots  \longrightarrow  
\frac{1}{4g_3^2} \tr F_{\mu \nu}\left[ 1   + 
{N_f  T(R)  \over 8} g_3^2  \frac{1}{ \sqrt \Box}   \right] F_{\mu \nu}  + \ldots
\label{RG0}
\end{equation}
where $T(R)$ is the index of the representation $R$,  the ellipsis stands for the fermionic terms, and the coefficient is the one appropriate to four-component complex Dirac fermions \cite{Appelquist:1989tc}. 
Defining a dimensionless coupling constant $\hat g_3^2 \equiv  {g_3^2  \over \mu}$, the one loop  renormalization group equation 
for  $\hat g_3^2 $ is given by:
\begin{equation}
\frac{d \hat g_3^2}{d \log \mu} =  -  \hat g_3^2\left(1 -  \frac{ N_f T(R)} {8}  \; \hat g_3^2 \right)~.
\label{RG1}
\end{equation}
Ref.~\cite{Appelquist:1989tc} pointed out the existence of a reliable conformal fixed point at large $N_f$ for fermions in the fundamental representation.  For the $ I=1/2 $ representation of  $SU(2)$,  we have $T(\half)=\half$ and indeed, it is necessary to take  $N_f$ large to  achieve a weakly coupled fixed point. 

Here,  we would like to point out an alternative class of conformal field theories which may be reached even by setting $N_f=1$. The idea is to use higher representations, i.e, to increase $T(R)$ rather than $N_f$, as both maneuvers increase the screening effects (the second term in (\ref{RG1})). 
 Of course, this would lead to the loss of asymptotic freedom on $\R^4$, but there are no such constraints on $\R^3$ as the theories of  interest have dimensionful couplings and the interactions are always damped at arbitrarily high energies  (the leading, classical running term in (\ref{RG1})).   
 
 At low energies,  the second term in (\ref{RG0}) dominates the gauge interactions. Hence, the 
 dimensionful $g_3^2$ drops out of dynamics, by an inverse dimensional transmutation. 
 Bringing the remaining gauge kinetic term into a ``canonical" form, we see that the dimensionless gauge coupling of CFT is $\hat g_{3,*}^2 \sim \frac{8}{T(R) N_f}$. At the fixed point, the photon propagator in momentum space is  $G(p) \sim {1\over p}$ and in real space, it is  
 $G(x) \sim {1\over x^2}$. The latter is the classical  propagator of gluons  in 4d as well. 
For $SU(2)$ gauge theory with $N_f=1$,  $T(j)= {1 \over 3}j (j+1) (2j+1)$. Thus, even with not so large $j$, one may achieve a weak coupling CFT with a fixed point 
$\hat g_{3,*}^2 \sim \frac{8}{T(j)}$.
Obviously, this class of CFTs may be easily generalized to $SU(N)$ gauge theories with higher-representation matter.

\section{An exotic magnetic bion on $\mathbf{\R^3}$}

Here we note that the fermionic mechanism, which renders the magnetic quintet in the chiral theory stable, has 
an elementary realization in three dimensions. 
 In \cite{Affleck:1982as}, Affleck, Harvey, and Witten discussed dynamical supersymmetry breaking and the role of instantons in $\R^3$.  
The type of topological operator  discussed in Section 3 of Ref.~\cite{Affleck:1982as}
did not find a physical interpretation so far. Here, we would like to fill this gap by showing how these operators explicitly arise.

The infrared physics of the model discussed in  Section 3 of  \cite{Affleck:1982as} is 
${\cal N}=1$ compact 
QED$_3$ with a single  real  ($d=3$) Majorana fermion. The instanton operator has the form:
\begin{equation}
{\cal M}= e^{-S_0} e^{i \sigma} \chi^T \gamma_0 \chi 
\label{monMaj}
\end{equation}
Under parity,   ${\cal P}: \chi^T \gamma_0 \chi \rightarrow - \chi^T \gamma_0 \chi$.  Since parity is a symmetry of the microscopic theory, the invariance of the instanton operator demands $\sigma \rightarrow \sigma  + \pi$ for the dual photon.  This means that the parity symmetry non-perturbatively intertwines  with the  $\Z_{2} \subset U(1)_J$ subgroup of  the topological shift symmetry 
of dual photon. Thus,
\begin{equation}
{\cal P} \times (\Z_2)_J \supset (\Z_2)_*
\end{equation}  
is  the symmetry that will be  used in the long-distance  effective theory. 
This means that  $e^{\pm i \sigma}$(or $\cos (\sigma)$)  is forbidden by the  $(\Z_2)_*$ symmetry, but  
 $e^{\pm 2 i \sigma}$ (or $\cos (2 \sigma)$) is not. Indeed,  the bosonic potential  derived from the superpotential also  leads to:
\begin{equation} 
{\cal B} + \overline {\cal B} =  2 e^{-2S_0} \cos (2 \sigma) \; .
 \label{bion}
\end{equation} In the fundamental domain $\sigma \in [0, 2\pi)$, the
theory has two isolated vacua and exhibits spontaneous parity breaking.
The topological excitation leading to the operator  $e^{\pm 2 i \sigma}$ remained elusive so far.
 
\begin{figure}[ht]
\begin{center}
\includegraphics[angle=0, width=0.4\textwidth]{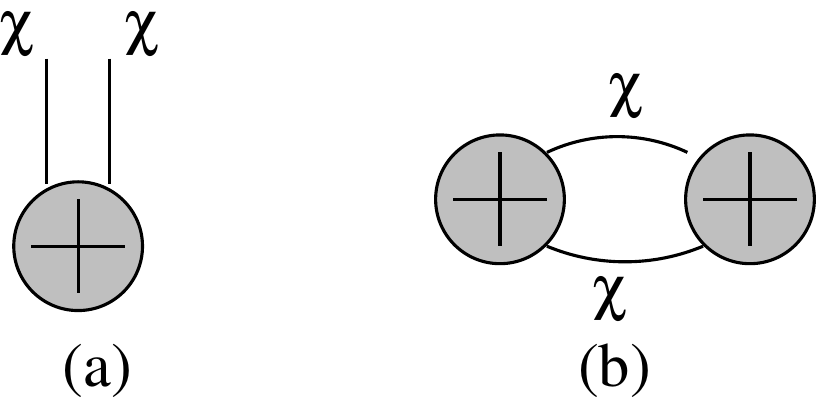}
\caption{
(a) is the  monopole operator ${\cal M}
$  (\ref{monMaj}) with two zero modes and charge normalized to $+1$. 
(b) is the magnetic bion operator (\ref{bion}), with no zero mode and charge $+2$. The latter is  stable via a fermionic paring mechanism, which overcomes Coulomb repulsion between constituent monopoles.  
 }
  \label {fig:monbion}
\end{center}
\end{figure}

It is evident that    $e^{2 i \sigma}$ has magnetic charge $+2$ and can be a composite of two charge $+1$ monopoles. However,  this immediately leads to  a puzzle, because of the $1/r$ Coulomb repulsion between constituents. This means that if such a term really exists in the Lagrangian---and by supersymmetry, we are certain that it does---there must be a mechanism which renders it stable. Other than gauge fluctuations, there is no {\it apparent} force carrier in the theory, and this constitutes a puzzle. How could there be such a bound state? To answer this question, 
 consider the   connected correlator  of two monopoles  evaluated in perturbative vacuum:
 \begin{eqnarray}
e^{-V(x-y)}= \langle {\cal M}(x){\cal M}(y)  \rangle_0 = e^{- \left(\frac{1}{|x-y|} + 2 \log|x-y|\right)} ~. 
 \end{eqnarray}
Apparently,  the fermion zero mode exchange 
of the Majorana fermions generates a logarithmically attractive interaction between two instantons of the same type, thus the topological excitation associated with the  $e^{2 i \sigma}$ 
operator is stable. 

 Such excitations were recently discovered in the context of 
both supersymmetric $\N\:$=$\:1$  and non-supersymmetric  gauge theories on  $S^1 \times \R^3$ and are referred to as magnetic bions \cite{Unsal:2007jx}.  For gauge theories on 
 $S^1 \times \R^3$, since the fermions are complex, the magnetically charged composite 
 is formed as  the bound state of 
${\cal M}_1= e^{i \sigma} \lambda \lambda$ (BPS) monopole and $\overline {\cal M}_2= e^{i \sigma} \bar \lambda \bar \lambda$ ($\overline {\rm KK}$) anti-monopole. 
  The magnetic bion in this case 
 is $ {\cal B} =  {\cal M}_1 \overline{\cal M}_2$.  Note that, with complex fermions, one cannot form an
  ${\cal M}_1 {\cal M}_1$ bound state, as 
 the net interaction between two such excitation  due to fermion zero mode exchange would vanish and no bound state would form. One can form an   ${\cal M}_1 \overline {\cal M}_1$ composite, but this is a dipole, which is magnetically neutral and   cannot generate a Debye mass for the photon.  
 
 When we reduce  $\N\:$=$\:1$ SYM or QCD with adjoint fermions down to $\R^3$, the analog of 
 the ${\cal M}_2$ topological excitation does not exist and the discrete axial symmetry is enhanced to a 
  $U(1)$ fermion number symmetry (a continuous  fermion number symmetry is absent for real Majorana fermions on $\R^3$ considered above). 
  Consequently, the intertwining of the continuous global symmetry with the topological $U(1)_J$ symmetry implies that  one can never form magnetic bions in 
   the ${\cal N}=2$ theory,  as well as in Polyakov models 
with complex adjoint Dirac fermions. This implies that theories with complex adjoint Dirac fermions 
in  $\R^3$  do not exhibit Abelian  confinement \cite{Affleck:1982as}, whereas the theory with real Majorana fermion does. The existence of a magnetic  bion in the latter is the main difference between these two classes of theories. We believe that  the presence or absence of the  magnetic bions explains all the interesting (unexplained) subtleties encountered in ref.~\cite{Affleck:1982as}.

\section{Monopoles of higher charge in valence-bond solid states of Heisenberg anti-ferromagnets}

Studies  \cite{Haldane:1988zz, Read:1990zza} of quantum anti-ferromagnets  on a square lattice 
have shown  that the  long distance dual formulation can be realized by  a Coulomb gas of monopoles with charges equal to multiples of the elementary monopole charge (we normalize the latter to 1), similar to the long-distance formulation of gauge theories with a center-stabilizing deformation. Consider an $SU(2)$  spin-$S$ anti-ferromagnet.  
Refs.~\cite{Haldane:1988zz, Read:1990zza} have shown that the leading monopole effects arising in the long distance effective theory are:
\begin{eqnarray}
e^{-QS_0} \cos (Q \sigma) ,~ Q=(1, 4,2, 4) \;    {\rm for} \;  2S=(0, 1,2, 3)\;  {\rm mod(4)} \;,
\label{HRS}
 \end{eqnarray}
causing  the appearance of $Q$ isolated vacua. Here, we would like to compare these topological excitations with the ones appearing in gauge theories, such as the magnetic bions described above and the magnetic quintets of the $I=3/2$ theory.

As described above,   in gauge theories  the monopole operator $e^{i \sigma}$ with charge $Q=1$ carries a certain number of fermionic zero modes dictated by the relevant  index theorem 
\cite{Poppitz:2008hr}. 
This leads to the intertwining of the discrete axial symmetry $(\Z_{h})_A$
with the topological shift symmetry $  (\Z_h)_J \subset U(1)_J$ such that:
\begin{equation}
(\Z_h)_A \times  (\Z_h)_J \supset (\Z_h)_*~,
\end{equation}  
is the realization of the symmetry in the long-distance topological operators. $(\Z_h)_*$ acts 
  on the dual photon as $\sigma \rightarrow \sigma+ \frac{2 \pi}{h}$.
Thus, the leading purely bosonic monopole operator is $e^{-h S_0} \cos(h \sigma)$.  This symmetry-based argument is backed-up by a dynamical ``pairing" mechanism, induced by 
multi-fermion exchanges, which renders these higher charge excitations stable. In gauge theories, the appearance of this class of purely bosonic operators at large distance implies the existence of $h$ isolated vacua.

The mechanism emerging in the Heisenberg anti-ferromagnets is equally interesting.
There, analogously, various monopole operators with $Q=1$ indeed exist. However, their fugacity picks a complex phase contribution which depends on $(2S)$ mod(4).  This is the crucial 
Berry phase as discovered in this context by Haldane  \cite{Haldane:1988zz}.  Soon after, 
Read and Sachdev wrote the long distance theory incorporating the Berry phase 
\cite{Read:1990zza}.   Remarkably, 
if   $(2S) \neq 0$ mod(4),  the Berry phase leads to destructive interference in the path integration, which leads  to the conclusion that the monopole operators with charge one drop out of dynamics. Furthermore, it also leads to a constructive interference which results in the monopole operator  given in (\ref{HRS}).

Thus, there are two mechanisms which render the composite topological excitations---with higher monopole charges, {\it e.g.}~(\ref{bion},\ref{HRS})---the leading effect in the long-distance  
non-perturbative dynamics. One is the Berry phase induced interference, and the other is the fermionic ``pairing" mechanism. The precise form of the composite topological excitation  in the latter is determined by the index theorem.  The relation  between the two mechanisms, if any, is currently unknown to us. However, the resulting topological excitations  play the same role in gauge theories and spin systems.

\acknowledgments
We  thank  M. Shifman, J. Harvey,  K. Intriligator, N. Seiberg, S. Shenker, P. Gao, and Y. Shang  for useful discussions.  This work was supported by the U.S.\ Department of Energy Grants DE-AC02-76SF00515 and by the National Science and Engineering Council of Canada (NSERC).




\bibliography{QCD}

\end{document}